# Quantifying the enhancement of two-photon absorption due to spectral-temporal entanglement

TIEMO LANDES,[1,2] MICHAEL G. RAYMER,[1,2] [*] MARKUS ALLGAIER,[1,2]
SOFIANE MERKOUCHE,[1,2] BRIAN J. SMITH,[1,2] AND ANDREW H. MARCUS[2,3]

[1] Department of Physics, University of Oregon, Eugene, OR 97403, USA
[2] Oregon Center for Optical, Molecular and Quantum Science, University of Oregon, Eugene, OR 97403, USA
[3] Dept of Chemistry and Biochemistry, University of Oregon, Eugene, OR 97403, USA
[*] raymer@uoregon.edu

**Abstract:** When a low flux of time-frequency-entangled photon pairs (EPP) illuminates a two-photon transition, the rate of two-photon absorption (TPA) can be enhanced considerably by the quantum nature of photon number correlations and frequency correlations. We present a quantum-theoretic derivation of entangled TPA (ETPA) and calculate an upper bound on the amount of quantum enhancement that is possible in such systems. The derived bounds indicate that in order to observe ETPA the experiments would need to operate at a combination of significantly higher rates of EPP illumination, molecular concentrations, and conventional TPA cross sections than are achieved in typical experiments.

## 1. Introduction

Photonic entanglement is key for many quantum applications spanning computing, communications, metrology and sensing. A large body of recent research in chemistry, physics and bioimaging explores the potential use of time-frequency-entangled photon pairs for applications that promise increased sensitivity at ultralow photon fluxes as well as increased simultaneous spectral and temporal resolving power. Many proposed applications rely on two-photon absorption (TPA) using entangled photon pairs, a process called entangled two-photon absorption (ETPA). But, in spite of the theoretical proposals, worldwide experimental efforts have yet to demonstrate convincing evidence of a quantum advantage of such techniques.

The potential 'quantum advantage' of entanglement in TPA spectroscopy poses a critical question that must be clarified in order for the field to move forward. This paper addresses the controversial role of time-frequency entanglement in two-photon molecular absorption, and provides a rigorous theoretical proof that the proposed methods are infeasible with technology in current use.

It is now widely accepted that when a low flux of time-frequency-entangled photon pairs (EPP) illuminates a two-photon transition, the rate of two-photon absorption (TPA) scales linearly with photon flux because the photons arrive in pairs, in contrast to the case of a coherent state wherein photons arrive randomly. And for a narrow TPA transition, photon frequency anticorrelation such as that generated in spontaneous parametric down conversion (SPDC) can lead to additional enhancement of the TPA probability if the sum of the paired photon frequencies sum to the frequency at the center of the TPA resonance.

Since the initial theoretical proposals by Gea-Banacloche, [[1]] and by Javanainen and Gould [[2]], followed by experimental confirmation for atomic systems by Georgiades et al [[3]] and by Dayan et al [4], these facts have entered into the lore of quantum optics, and more recently quantum information science where 'quantum advantage' is sought for creating capabilities not possible when using 'classical' states of light. A persistent goal has been to apply TPA with EPP, referred to as entangled TPA (or ETPA) to molecular systems in solution or in biological systems. A wide range of proposals for enhancing spectroscopy have been made, [5, 6] and several suggestive experiments have been carried out. [7, 8]

The key question that has not yet been answered definitively is—are ETPA events frequent enough in typical molecular systems for them to be detectable using current technology? Recent experiments called into question the observability of TPA driven by isolated EPP, where 'isolated' indicates the regime of extremely low flux where not more than two photons on average impinge on the molecule within the field's coherence time and not more than two photons impinge on the molecule within its coherent response time.

We first review the standard predictions and state the problem quantitatively. We then demonstrate theoretically that, according to this calculation, the answer to the key question is, "ETPA events are not frequent enough to produce detectable signals in typical molecular systems using currently-applied SPDC sources." This conclusion is consistent with recent experimental results both in the isolated EPP regime [9,10] and at higher fluxes. [11]

Fei et al [12] developed an often-cited model for ETPA, in which the two-photon absorption rate for a molecule illuminated at very low flux by isolated time-frequency-entangled photon pairs (EPP) is given by $R = \sigma_e I$, where $I$ is the photon flux density (with units $m^{-2}s^{-1}$) and $\sigma_e$ is generally called the entangled two-photon absorption cross section (with units $m^2$). [13] This linear scaling of rate with flux density was first pointed out in [1, 2]. Fei et al proposed a heuristic model for ETPA, which can be summarized by:

$$\sigma_e = \frac{\sigma^{(2)}}{A_e T_e} \times f_{EPP} \qquad (1)$$

where $\sigma^{(2)}$ is the conventional two-photon absorption cross section first derived by Maria Göppert-Mayer [14]. For molecules in solution $\sigma^{(2)}$ is typically very small—of the order of 1 to 1,000 GM (where $1 GM = 10^{-58} m^4 s$). [15] $A_e$ is the 'entanglement area,' that is the area within which EPP are spatially correlated. This area may be smaller than the area over which EPP are spread. In experiments where the EPP are tightly focused into a molecular sample $A_e$ is roughly equal to or slightly smaller than the focal spot area, so that in this case no significant quantum enhancement results from this factor. $T_e$ is the 'entanglement time,' that is the interval within which EPP are temporally correlated. This time may be shorter than the interval over which EPP are spread. $T_e$ equals roughly the inverse of the optical bandwidth of the EPP beam and can lead to considerable enhancement of the ETPA rate at low flux. [16] $f_{EPP}$ is a unitless factor, whose magnitude depends on the nature of the entangled state of the EPP and the parameters of the two-photon transition.

A controversy exists over the magnitude of $f_{EPP}$, and thus of the ratio of entangled-to conventional cross sections $\sigma_e / \sigma^{(2)}$. Several experimental studies have indicated that ETPA can be readily detected using 10 orders of magnitude fewer photons than required when using conventional light source that are not entangled. [17] Other experimental studies place the ETPA cross section many orders of magnitude smaller. [9, 11] Without the presence of extremely large values of $\sigma_e / \sigma^{(2)}$, two-photon absorption at very low flux levels, where the quantum advantage is expected to play a role, would not be observable in typical experiments, due to the extremely small values of $\sigma^{(2)}$, as illustrated below.

In this paper we show theoretically that in cases in which ETPA occurs only via far-off-resonant (virtual) intermediate states, such that no appreciable population in or mutual coherence between intermediate states exists, the value of $f_{EPP}$ is bounded by a value of order unity. Therefore, the extremely large quantum enhancement that would be needed to observe TPA signals at low photon flux is not present in molecules with typical two-photon cross-sections.

## 2. Theory of ETPA

We express the above-discussed predictions in terms of a single pulse of EPP incident on a molecule. If the rate expression is integrated over the duration of the pulse, and assuming the entanglement area equals the beam area $A_0$, the probability for the molecule to transition from the ground state to the final state $f$ can be written as:

$$P_f = \frac{N_{EPP}}{A_0}\sigma_e = \left(\frac{N_{EPP}}{A_0}\right)\frac{\sigma^{(2)}}{A_0 T_e} \times f_{EPP} \qquad (2)$$

where $N_{EPP} < 1$ is the mean number of photons in the EPP pulse of duration $T_p$.

A rigorous expression for the probability for two-photon absorption (TPA), valid to fourth-order in interaction strength has been derived in numerous publications, but to date has not been analyzed in terms of its detectability in typical molecules. [1, 2, 5, [18]] When intermediate-state populations and mutual coherences can be neglected and TPA proceeds only through coherent pathways (off-diagonal in the density matrix formulation), the probability for a molecule to transition from the ground state $g$ to the final state $f$ by a pulsed field described by any pure state $|\Psi\rangle$ can be expressed as: [[19]]

$$P_f = \frac{\sigma^{(2)}}{A_0^2} \int \dd\omega \int \dd\tilde{\omega} \int \dd\omega' \, \mathcal{L}(\omega_{fg} - \omega - \tilde{\omega}) C(\omega, \tilde{\omega}, \omega') \qquad (3)$$

where we use the notation $\dd\omega = d\omega/2\pi$. The field correlation function in the frequency domain is:

$$C(\omega, \tilde{\omega}, \omega') = \langle \Psi | \hat{a}^\dagger(\omega') \hat{a}^\dagger(\omega + \tilde{\omega} - \omega') \hat{a}(\omega) \hat{a}(\tilde{\omega}) | \Psi \rangle \qquad (4)$$

with $\hat{a}(\omega)$ being the annihilation operator for a photon with (angular) frequency $\omega$, obeying $[\hat{a}(\omega'), \hat{a}^\dagger(\omega)] = 2\pi\delta(\omega' - \omega)$. We assume the TPA transition in the molecule is broadened homogeneously by dephasing interactions with an environment leading to a half linewidth $\gamma_{fg}$ and a Lorentzian line shape, peak-normalized ($\mathcal{L}(0) = 1$):

$$\mathcal{L}(\omega_{fg} - \omega - \tilde{\omega}) = \frac{\gamma_{fg}^2}{\gamma_{fg}^2 + (\omega_{fg} - \omega - \tilde{\omega})^2} \qquad (5)$$

The TPA cross section that emerges from the quantum treatments is [19]:

$$\sigma^{(2)} = \left(\frac{\hbar\omega_0}{\varepsilon_0 nc}\right)^2 \frac{1}{2\gamma_{fg}} \sum_{m,m'} \frac{\mu_{fm}\mu_{mg}\mu_{m'f}\mu_{gm'}}{(-\omega_{fm}+\omega_0)(\omega_{m'g}-\omega_0)} \quad (6)$$

where $\omega_0$ is the central frequency of the exciting light, $\omega_i$ are the frequencies of the states labeled $i$, and $\mu_{ij} = \mathbf{d}_{ij} \cdot \mathbf{e}/\hbar$ are the (scaled) electric-dipole transition matrix elements projected onto the polarization $\mathbf{e}$ of the field. $A_0$ is the effective beam area at the molecule's location, $n$ is the refractive index of the medium, $\varepsilon_0$ is the vacuum permittivity and $c$ is the vacuum speed of light. Equation (6) for the cross section agrees precisely with the conventional one derived using second-order perturbation theory and averaging over the density of states, and for molecules in solution $\sigma^{(2)}$ is typically of the order of 1 to 1,000 GM (where $1\,GM = 10^{-58}\,m^4 s$). [15] Here the field polarization is taken to be common to all the incident light, that is, the field is in a single spatial-polarization mode, as is produced by forward colinear Type-I spontaneous parametric down conversion (SPDC). If Type-II SPDC is used instead, the formulas are modified slightly, and the conclusions are not altered.

Collinear Type-0 or Type-I spontaneous parametric down conversion pumped by a pulse of finite duration can be designed to occur into a single spatial-and-polarization mode [20]; then the state is described by:

$$|\Psi\rangle = \sqrt{1-\varepsilon^2}|vac\rangle + \varepsilon \int d\omega \int d\tilde{\omega}\, \psi(\omega,\tilde{\omega})\hat{a}^\dagger(\omega)\hat{a}^\dagger(\tilde{\omega})|vac\rangle \quad (7)$$

The probability that a given pulse contains a photon pair is $\varepsilon^2 \ll 1$. We neglect higher-order terms representing generation of multiple pairs in order to satisfy our assumption of isolated EPP interacting with the molecule. The joint-spectral amplitude (JSA) $\psi(\omega,\tilde{\omega})$ is determined by the spectrum of the pumping laser pulse and the phase-matching properties of the nonlinear crystal used as the second-order nonlinear medium. [21] It is normalized: $\int d\omega \int d\tilde{\omega}\, |\psi(\omega,\tilde{\omega})|^2 = 1$. For Type-0 or Type-I SPDC the JSA is symmetric, $\psi(\omega,\tilde{\omega}) = \psi(\tilde{\omega},\omega)$. The mean numbers of photons (twice the pairs number) in the pulse is $N_{EPP} = 2\varepsilon^2$.

The field correlation function is then found to equal:

$$C(\omega,\tilde{\omega},\omega') = 4\varepsilon^2 \psi^*(\omega',\omega+\tilde{\omega}-\omega')\psi(\omega,\tilde{\omega}) \quad (8)$$

In the case of Type-0 or Type-I SPDC with the two photons occupying distinct spatial modes, or for Type-II SPDC, where the signal and idler modes occupy distinct polarization modes, Eq.(8) still holds, with the replacement: $\psi(\omega,\tilde{\omega}) \to \Psi(\omega,\tilde{\omega}) = \{\psi(\omega,\tilde{\omega})+\psi(\tilde{\omega},\omega)\}/2$. Thus, the relevant quantity is the 'two-photon amplitude' $\Psi(\omega,\tilde{\omega})$, which satisfies exchange symmetry even though the JSA $\psi(\omega,\tilde{\omega})$ is not necessarily symmetric. [19]

The probability is maximum when twice the center frequency of the EPP equals the TPA transition frequency, that is $2\omega_0 = \omega_{fg}$. Taking that to be the case, then by a change of variables, $x = (\omega + \tilde{\omega} - 2\omega_0)$, $z = \omega - \omega_0$, and inserting Eq.(8), we can write Eq.(3) as:

$$P_f = 2\left(\frac{N_{EPP}}{A_0}\right)\frac{\sigma^{(2)}}{A_0}\int dx\, \mathcal{L}(x)|K_\psi(x)|^2 \qquad (9)$$

where in the general case:

$$K_\psi(x) = \int dz\, \Psi(\omega_0 + z, \omega_0 + x - z) \qquad (10)$$

is the integrated amplitude (antidiagonal projection) for TPA at a particular two-photon detuning *x*. A graphical interpretation is given in **Fig. 1**.

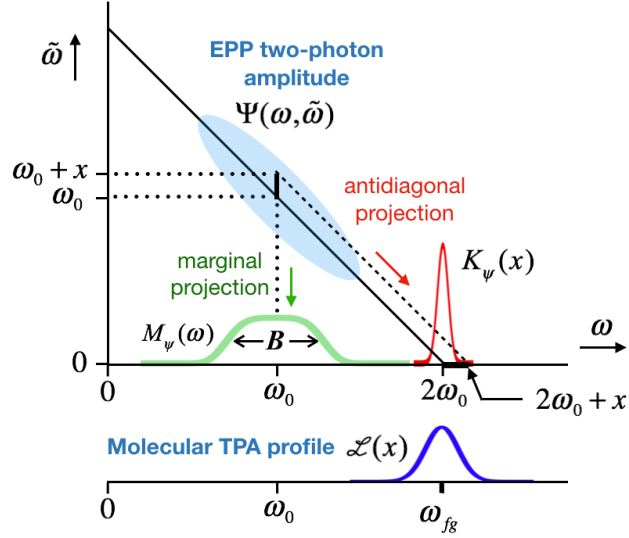

Fig. 1 The antidiagonal projection $K_\psi(\omega)$ of the two-photon amplitude $\Psi(\omega, \tilde{\omega})$, and the two-photon absorption profile $\mathcal{L}(x)$, assumed to be two-photon resonant with the center frequency of the EPP light field. Also shown is the 'marginal' projection, $M_\psi(\omega) = \int d\tilde{\omega}\,|\Psi(\omega, \tilde{\omega})|^2$, which is the energy spectrum of the EPP and has bandwidth *B*.

We write Eq.(9) as:

$$P_f = \left(\frac{N_{EPP}}{A_0}\right)\frac{\sigma^{(2)}}{A_0}\eta \qquad (11)$$

where we defined the 'spectral overlap factor' as:

$$\eta \equiv 2\int dx\, \mathcal{L}(x)|K_\psi(x)|^2 \qquad (12)$$

Comparing Eq.(11) to Eq.(2), we can identify $\eta = f_{EPP}/T_e$, and the question is whether $\eta$ can be many orders of magnitude greater than the inverse of the entanglement time $T_e$.

The value of $\eta$ can be upper-bounded and compared to the entanglement time as follows. The line shape $\mathcal{L}(x)$ is a property of the given molecule and thus for our purposes cannot be varied.

It is clear from the projection geometry in Fig. 1 that the form of $|K_\psi(x)|^2$ that maximizes $\eta$ is that leading to the greatest amount of spectral compression of the two-photon amplitude into the TPA absorption line $\mathcal{L}(x)$. This situation occurs when $|K_\psi(x)|^2$ is narrow along the diagonal and broad along the antidiagonal, that is, the entanglement is greatest. Such a form occurs naturally in SPDC when pumped by a laser pulse that is long compared to the inverse phase-matching linewidth. Thus, to find an expression for the upper bound of the integral, we assume the two-photon amplitude can be written in factored form as the product of narrow and broad functions, $\psi_N(\omega)$ and $\psi_B(\omega)$ respectively, centered at $\omega = \tilde{\omega} = \omega_0$ and oriented along diagonal and antidiagonal axes in the $(\omega, \tilde{\omega})$ plane respectively:

$$\Psi(\omega, \tilde{\omega}) = \psi_N(\omega + \tilde{\omega} - 2\omega_0)\psi_B(\frac{\omega - \tilde{\omega}}{2}) \qquad (13)$$

where, as required by state symmetry, $\psi_B(-x) = \psi_B(x)$, and both functions are square-normalized in $đ\omega = d\omega/2\pi$. The spectral width of $\psi_N(\omega)$ is the linewidth (inverse duration) of the pump pulse, and the width of $\psi_B(\omega)$ is determined by the phase-matching in the SPDC source crystal. The ratio of these two widths is a measure of the amount of time-frequency entanglement in the EPP field. [22] The more elongated the two-photon amplitude is, the higher the degree of entanglement. This model can be a good approximation for Type-0, -I, or -II SPDC, depending on pulse durations and phase-matching conditions. [23] Note also that under the assumption that the state is highly entangled, the marginal distribution is given approximately by the square of the broad function, $M_\psi(\omega) \approx |\psi_B(\omega - \omega_0)|^2$.

The entanglement time can be evaluated under the optimal factorization assumption. Defining difference and sum times as $\tau = t - \tilde{t}$, $\tilde{\tau} = (t + \tilde{t})/2$, double Fourier transform of Eq.(13) gives the two-photon amplitude in the time domain as:

$$\begin{aligned}\Phi(t,\tilde{t}) &= \int đ\omega \int đ\tilde{\omega}\, \Psi(\omega, \tilde{\omega}) e^{-i\omega(\tilde{\tau}+\tau/2)} e^{-i\tilde{\omega}(\tilde{\tau}-\tau/2)} \\ &= \varphi_N(\tilde{\tau})\varphi_B(\tau)\end{aligned} \qquad (14)$$

and $\varphi_N(\tilde{\tau})$, $\varphi_B(\tau)$ are Fourier transforms of $\psi_N(\omega)$ and $\psi_B(\omega)$, respectively. Using a change of variables, $y = \omega + \tilde{\omega}, z = (\omega - \tilde{\omega})/2$, we find:

$$\begin{aligned}\varphi_N(\tilde{\tau}) &= \int đy\, \psi_N(y - 2\omega_0) e^{-iy\tilde{\tau}} \\ \varphi_B(\tau) &= \int đz\, \psi_B(z) e^{-iz\tau}\end{aligned} \qquad (15)$$

The entanglement time $T_e$ is defined as the 'width' of the difference-time distribution $|\varphi_B(\tau)|^2$, as illustrated in **Fig. 2**. We will quantify 'width' in discussion below.

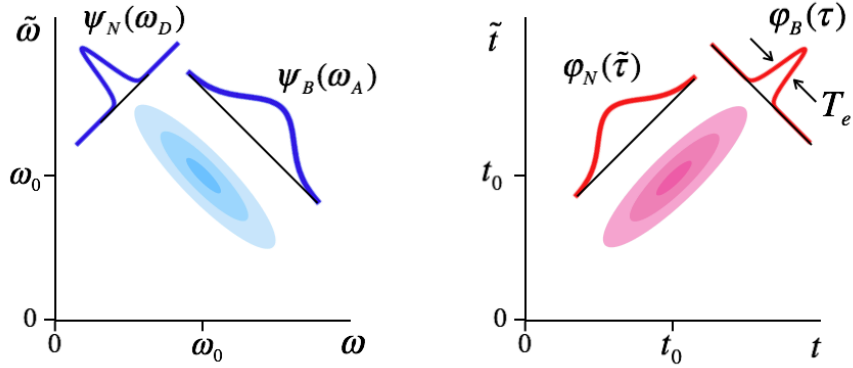

**Fig. 2**. Two-photon amplitude in frequency domain $\Psi(\omega,\tilde{\omega})$ and time domain $\Phi(t,\tilde{t})$. Diagonal and antidiagonal frequency arguments are $\omega_D = \omega + \tilde{\omega} - 2\omega_0$ and $\omega_A = (\omega - \tilde{\omega})/2$. Diagonal and antidiagonal time arguments are $\tau = t - \tilde{t}$ and $\tilde{\tau} = (t + \tilde{t})/2$.

Now we note that Eq.(13) leads to:

$$K_\psi(x) = \psi_N(x) \int dz\, \psi_B(z - x/2) \qquad (16)$$

Thereby, Eq.(12) equals the product of two factors, $\eta \equiv 2\eta_N \eta_B$, where:

$$\begin{aligned}\eta_N &= \int dx\, \mathcal{L}(x) |\psi_N(x)|^2 \\ \eta_B &= \left|\int dz\, \psi_B(z)\right|^2\end{aligned} \qquad (17)$$

Upper bounds on the two factors can be determined as follows. Noting that $\mathcal{L}(x)$ is everywhere less than or equal to 1, we conclude that $\eta_N$ is maximum when $|\psi_N(x)|^2$ is much narrower than $\mathcal{L}(x)$, yielding the greatest amount of spectral compression of the EPP spectrum into the TPA line. Then:

$$\eta_N^{max} = \int dx\, |\psi_N(x)|^2 = 1 \qquad (18)$$

The other factor, $\eta_B$, is maximized when $\psi_B(z)$ is constant over the region in which it is defined, as shown in the **Appendix** via the Cauchy-Schwartz inequality. Note that the integration variable $x = (\omega + \tilde{\omega} - 2\omega_0)$ is restricted to the range $[0, 2\omega_0]$ because the spectrum of EPP generated by SPDC with a pump laser with frequency $2\omega_0$ cannot extend above $2\omega_0$. Therefore, we can consider $\psi_B(\omega)$ to be defined and normalized on the finite region $[\omega_0 - \Omega, \omega_0 + \Omega]$, for which $\eta_B$ is maximized by $\psi_B(z) = (\pi/\Omega)^{1/2}$, giving:

$$\eta_B{}^{max} = \left| \int_{\omega_0-\Omega}^{\omega_0+\Omega} \sqrt{\frac{\pi}{\Omega}} dz \right|^2 = \frac{\Omega}{\pi} \qquad (19)$$

Under the factorization assumption in Eq.(13), which is optimal, and the two bounds derived above, the maximum possible value of $\eta \equiv 2\eta_N \eta_B$, for a given maximum bandwidth $2\Omega$ of the EPP field, is:

$$\eta^{max} = 2\eta_N{}^{max}\eta_B{}^{max} = \frac{2\Omega}{\pi} \qquad (20)$$

To interpret this result, note also that $\eta_B$ is equal to the Fourier time-domain function $\varphi_B(\tau)$ evaluated at $\tau = 0$:

$$\eta_B = \left| \int dz\, \psi_B(z) \right|^2 = |\varphi_B(\tau = 0)|^2 \qquad (21)$$

For the optimal case, $\psi_B(z) = (\pi/\Omega)^{1/2}$, we find for the time-domain function:

$$\varphi_B(\tau) = \frac{\Omega^{1/2}}{\pi^{1/2}} e^{-i\omega_0 \tau} \left( \frac{\sin(\Omega \tau)}{\Omega \tau} \right) \qquad (22)$$

which is consistent with Eqs.(21) and (19) for $\tau = 0$. It is understandable that the maximum ETPA occurs if the photon pairs are correlated in time as tightly as possible, making $|\varphi_B(\tau = 0)|^2$ large.

As described above, we can reconcile Eqs.(2) and (11) quantitatively by equating:

$$\eta \equiv \frac{f_{EPP}}{T_e} \qquad (23)$$

The precise value of $f_{EPP}$ will depend on how the entanglement time $T_e$ is quantified, which is dependent on the chosen convention. For the case of interest, where ETPA is maximized, $\varphi_B(\tau)$ is a simple, smooth function peaked around $\tau = 0$. In this case we can use a convenient approximation for the bandwidth $B$ and temporal duration $T$ of a Fourier-transform pair of functions $f(\omega)$ and $\tilde{f}(t)$ (both square-normalized), as discussed in the **Appendix**. This convention leads to:

$$\begin{aligned} B &= \int \frac{d\omega}{2\pi} \frac{|f(\omega)|^2}{|f(\omega_{max})|^2} = \frac{1}{|f(\omega_{max})|^2} \\ T &= \int dt\, \frac{|\tilde{f}(t)|^2}{|\tilde{f}(t_{max})|^2} = \frac{1}{|\tilde{f}(t_{max})|^2} \end{aligned} \qquad (24)$$

These definitions of entanglement time and bandwidth yield the bound:

$$\eta_B = |\varphi_B(\tau = 0)|^2 \frac{|\varphi_B(\tau_{max})|^2}{|\varphi_B(\tau_{max})|^2} = \frac{1}{T_e} \frac{|\varphi_B(\tau = 0)|^2}{|\varphi_B(\tau_{max})|^2} \leq \frac{1}{T_e} \qquad (25)$$

For the case where the overall ETPA probability is maximized, $f_{EPP}$ is bounded by: $f_{EPP} \equiv \eta T_e = \eta / \eta_B = 2\eta_N = 2$. In cases where overall ETPA is not maximized, it may in principle be possible to engineer complex temporal functions for which $f_{EPP}$ is large for a given definition of $T_e$. But this scenario cannot increase the overall ETPA efficiency over the demonstrated bound.

In practice, the entanglement time will be larger than its lower bound $T_e \simeq 2\pi/\Omega$, and can be conveniently approximated as the inverse of the marginal bandwidth, $B$, of the EPP spectrum: $T_e \approx 1/B$. Thus, the quantum-enhanced TPA cross section will typically have values significantly less than $\sigma_e = \sigma^{(2)}\Omega/\pi A_0$, and instead will be:

$$\sigma_e \simeq \frac{\sigma^{(2)}}{A_0} \frac{B}{2\pi} \qquad (26)$$

where $B/2\pi$ is the bandwidth of the EPP spectrum in units of $Hz$. The final-state probability will then be:

$$P_f \simeq \left(\frac{N_{EPP}}{A_0}\right) \frac{\sigma^{(2)}}{A_0} \frac{B}{2\pi} \qquad (27)$$

## 3. Experimental feasibility of observing ETPA

Consider an EPP beam centered at 1064 *nm* with 40 *nm* marginal bandwidth, focused to a waist radius of 5 *um* into a 1 *cm*-long cuvette of Rhodamine 6G dye solution which has a two-photon absorption cross-section of 9 GM. In this case the effective beam area is ~8x10$^{-7}$ $cm^2$ and the effective focal volume is about 1.2x10$^{-8}$ $cm^3$. There are approximately 7.0x10$^9$ molecules in this volume per mmol concentration. The bandwidth, B/2$\pi$, is about 1.1x10$^{13}$ *Hz*. The probability of a single R6G molecule absorbing an EPP with these characteristics can be estimated using Eq. 28, which yields $P_f = 1.5 \times 10^{-24}$. And the probability of an EPP being absorbed in the solution is 1.1x10$^{-14}$ per mmol, which is far too small a fractional absorbance to be measured via transmittance, regardless of the concentration or EPP rate.

Background-free measurements such as fluorescence are more sensitive to changes of this magnitude. Consider a pulsed laser system with a repetition rate of 80 MHz that produces SPDC photons with probability $\varepsilon^2 = 0.1$. The expected rate of TPA events in a 100 *mmol* solution of Rhodamine 6G is 8.8x10$^{-6}$ $s^{-1}$ for such a system. This is again far below any reasonable detection thresholds. For CW excitation schemes, the EPP rate can, in principle, be increased to around $10^{13}$ $s^{-1}$ for EPP with these characteristics before temporal overlap between pairs dominates, invalidating the isolated-EPP condition. In this case, the expected rate of absorption in a 100 *mmol* solution of R6G is 11 $s^{-1}$.

While this rate is in-principle detectable via background-free fluorescence measurements using standard detectors, additional considerations such as the quantum efficiency of the fluorophore (*QE*), the collection efficiency of the experimental apparatus (*CE*), the detector efficiency, and the detector dark rate, make detecting fluorescence of this magnitude highly problematic. In particular, the fluorescence efficiency for solutions that are so highly concentrated is known to suffer from self-quenching, reabsorption, and other effects that reduce the amount of detectable fluorescence. Thus, detecting ETPA with spectroscopically relevant concentrations (typically micro-molar) appears to be infeasible.

For a beam of similar spatial profile, and an arbitrary molecule, we can write the expected rate using intuitive units: 1.2x10$^{-15}$ /GM/mmol/pair. A system capable of producing a combined value of *GM* x *mmol* x *pairs* x *QE* x *CE* >> 10$^{15}$ s$^{-1}$ would in-principle be capable of detecting ETPA with fluorescence.

The theory presented in this paper indicates that observing ETPA in molecules without a resonant intermediate state is infeasible without simultaneously, ultra-high EPP flux, ultra-high concentrations, and ultra-large conventional two-photon absorption cross-sections. Observations of absorbed EPP in experiments that do not meet these conditions, could point to other effects in the solution causing loss or scatter, that are not accounted for by non-resonant TPA.

## 4. Comparison of EPP and coherent-state TPA

We have verified by straightforward calculation, for cases where intermediate-state populations and mutual coherences do not play a significant role in ETPA, that any enhancement resulting from the spectral properties of the EPP field arises only from the distinction between the entanglement time $T_e \approx 1/B$ and the total duration of the EPP pulse, which may be much longer than $T_e$.

To verify this statement, compare ETPA to TPA driven by a coherent-state pulse of duration $T_c$, following which the excited-state population is defined as:

$$P_f^{coh} \approx \left( \frac{N_{coh}^2}{A_0} \right)\left( \frac{\sigma^{(2)}}{A_0 T_c} \right) \times f_{coh} \qquad (28)$$

where $N_{coh}$ is the mean number of photons in the pulse. By a calculation analogous to the one above, the factor $f^{coh}$ is found to be bounded by a value of order unity, the exact value depending on the pulse shape. [19] The enhancement in the rate of TPA by EPP in comparison to excitation by coherent-state light can be characterized by the 'quantum enhancement factor'(QEF), defined as the ratio of excited populations in the two cases. In many cases of interest for this comparison one can show that $f^{coh} \simeq f_{EPP}$. This equality holds when the anti-diagonal projection of the joint spectral amplitudes of the EPP and the coherent-state comparator pulse are equal, or both are much narrower than the linewidth of the two-photon transition. In these cases, the QEF can be written as:

$$QEF = \frac{P_f^{EPP}}{P_f^{coh}} = \left( \frac{N_{EPP}}{N_{coh}^2} \right)\left( \frac{T_c}{T_e} \right) \qquad (29)$$

In the regime of isolated (nonoverlapping) EPP we have $N_{EPP} < 1$. Then, for comparison, if the two pulses have the same mean photon number, $N_{EPP} = N_{coh} = N$, we have:

$$QEF = \left(\frac{1}{N}\right)\left(\frac{T_c}{T_e}\right) \qquad (30)$$

While decreasing the value of *N* does result in an increase in QEF, the net result of decreasing *N* will always be a reduction in overall signal intensity.

## 5. Conclusions

We presented a quantum-theoretic derivation of time-frequency entangled two-photon absorption (ETPA) and calculated an upper bound on the amount of quantum enhancement made possible by photon number correlations and frequency anticorrelations. Our theory focuses on the off-resonant two-photon excitation pathway that goes through (off-diagonal) dipole coherences rather than pathways that go through populations and mutual coherences of intermediate states, often referred to as rephasing and non-rephasing pathways. Thus, our treatment is valid when there are no intermediate states whose transitions overlap spectrally with the exciting light. Such a case occurs, for example, if all intermediate states lie above the energy of the final state excited in the two-photon transition. While the enhancement is found to be significant if the EPP are highly entangled spectrally, the theory predicts that in the isolated-pair regime ETPA events are not frequent enough in typical molecular systems to be detectable in realistic experiments. We predict this to be the case for both fluorescence-detected and transmission-detected TPA.

The situation may be different if the molecule has intermediate states that lie well below the energy of the final state excited in the two-photon transition and overlap spectrally the EPP spectrum, such that a stepwise sequence of excitations through intermediate-state populations and mutual coherences is possible. However, it is expected that in such cases the advantage of spectral anticorrelations will be partly or completely lost. [16] A factor that would complicate such experiments on fluorescence-detected molecular TPA is the presence of one-photon-excited background signals, since nonradiative decay typically is rapid, leading to the fluorescence signal arising from states that can also be excited by one-photon absorption.

To observe quantum enhancement of TPA by photon entanglement in the isolated-pair regime, one should use systems with narrow-band transitions, such as atoms or individual color centers in crystals, such that the frequency anticorrelations in the EPP would be most effective. [4] In such cases, ETPA may prove to be useful a method for enhancing spectroscopy.

**Acknowledgements**

This work was supported by grants from the National Science Foundation RAISE-TAQS Program (PHY-1839216 to M.G.R., A.H.M., and B.S. as co-PIs).

**Data Availability**

The data supporting this study are contained within the article.

**Disclosures**

The authors declare no conflicts of interest.

**Appendix**
**A1. Cauchy-Schwarz Inequality**

For square-integrable functions defined on a domain, D, with the inner-product norm, the Cauchy-Schwartz inequality holds and can be stated as:

$$\left| \int_D dx\, g(x) h^*(x) \right|^2 \leq \int_D |g(x)|^2 dx \int_D |h(x)|^2 dx \qquad (31)$$

The inequality is saturated when $g(x) = \lambda h(x)$ for complex $\lambda$. This inequality can be applied straightforwardly to set a bound on $\eta_B$ by noting that $\psi_B(\omega)$ is square-normalized and necessarily restricted to a finite frequency range $[0, 2\omega_0]$ for SPDC entangled photons. If we consider some range, $[\omega_0 - \Omega, \omega_0 + \Omega]$ with $g(z) = 1$, then (from Eq.(17):

$$\eta_B = \left| \int \frac{dz}{2\pi} g(z) \psi_B(z) \right|^2 \leq \int_D \frac{dz}{2\pi} |g(z)|^2 \cdot \int_D \frac{dz}{2\pi} |\psi_B(z)|^2 \qquad (32)$$

This inequality is saturated when $\psi_B(z) = \lambda g(z) = \lambda$, with $\lambda^2 = \pi/\Omega$ to ensure $\psi_B(z)$ is square normalized.

**A2. Bandwidth and duration estimates for square-normalized functions**

Here we present a formalism for defining the spectral bandwidth $B$ and temporal duration $T$ of Fourier-transform pairs. Given a real, positive-definite function of angular frequency, $F(\omega)$, whose maximum value equals one, $F(\omega_{max}) = 1$ we define its full bandwidth $B$ (in units of *Hz*) as:

$$B \equiv \int \frac{d\omega}{2\pi} F(\omega) \qquad (33)$$

The bandwidth in units of *rad/s* is $2\pi B$. And given a real, positive-definite function of time, $G(t)$, whose maximum value equals one, $G(\omega_{max}) = 1$ we define its full duration $T$ (units of seconds) as:

$$T \equiv \int dt\, G(t) \qquad (34)$$

These definitions have been used successfully in studies of spectral and temporal filtering of quantum light. [24, 25] They provide general approximations to the full widths for simple peak-normalized functions such as Gaussians, Lorentzians, and sinc-squared, as illustrated for duration in **Table 1**.

**Table 1**. Examples of duration as defined by Eq.(34)

| Peak-normalized function | Full width half max | Duration |
|---|---|---|
| $\Pi(x/\sigma)$ * | $\sigma$ | $\sigma$ |
| $\exp(-x^2/2\sigma^2)$ | $2(2\ln 2)^{1/2} = 2.35\sigma$ | $(2\pi)^{1/2}\sigma = 2.51\sigma$ |
| $\text{sinc}^2(x/\sigma)$ | $2.78\sigma$ | $\pi$ |
| $\sigma^2/(\sigma^2 + x^2)$ | $2\sigma$ | $\pi\sigma$ |

* Heaviside Pi (Box function)

Now, consider a square-normalized Fourier transform pair:

$$\tilde{f}(t) = \int d\omega\, f(\omega) e^{-i\omega t} \quad , \quad f(\omega) = \int dt\, \tilde{f}(t) e^{i\omega t} \qquad (35)$$

It follows directly that:

$$B = \int \frac{d\omega}{2\pi} \frac{|f(\omega)|^2}{|f(\omega_{max})|^2} = \frac{1}{|f(\omega_{max})|^2}$$

$$T = \int dt \frac{|\tilde{f}(t)|^2}{|\tilde{f}(t_{max})|^2} = \frac{1}{|\tilde{f}(t_{max})|^2} \qquad (36)$$

## References

[1] Julio Gea-Banacloche, "Two-photon absorption of nonclassical light," *Physical review letters* **62**, 1603 (1989).

[2] Juha Javanainen and Phillip L. Gould. "Linear intensity dependence of a two-photon transition rate." *Physical Review A* **41**, no. 9 5088 (1990).

[3] N. Ph Georgiades, E. S. Polzik, K. Edamatsu, H. J. Kimble, and A. S. Parkins. "Nonclassical absorption for atoms in a squeezed vacuum." *Physical review letters* **75**, no. 19, 3426 (1995).

[4] Barak Dayan, Avi Pe'er, Asher A. Friesem, and Yaron Silberberg. "Two photon absorption and coherent control with broadband down-converted light." *Physical review letters* **93**, no. 2 (2004): 023005.

[5] Frank Schlawin, Konstantin E. Dorfman, and Shaul Mukamel. "Entangled two-photon absorption spectroscopy." *Accounts of chemical research* **51**, no. 9 (2018): 2207-2214.

[6] Michael G. Raymer, Andrew H. Marcus, Julia R. Widom, and Dashiell LP Vitullo. "Entangled photon-pair two-dimensional fluorescence spectroscopy (EPP-2DFS)." *The Journal of Physical Chemistry B* **117**, no. 49 (2013): 15559-15575.

[7] Dong-Ik Lee and Theodore Goodson. "Entangled photon absorption in an organic porphyrin dendrimer." *The Journal of Physical Chemistry B* **110**, no. 51 (2006): 25582-25585.

[8] Juan P. Villabona-Monsalve, Omar Calderón-Losada, M. Nuñez Portela, and Alejandra Valencia. "Entangled two photon absorption cross section on the 808 nm region for the common dyes zinc tetraphenylporphyrin and rhodamine b." *The Journal of Physical Chemistry A* **121**, no. 41 (2017): 7869-7875.

[9] Tiemo Landes, Markus Allgaier, Sofiane Merkouche, Brian J. Smith, Andrew H. Marcus, and Michael G. Raymer. "Experimental feasibility of molecular two-photon absorption with isolated time-frequency-entangled photon pairs." *arXiv preprint arXiv:2012.06736* (2020).

[10] Samuel Corona-Aquino, Omar Calderón-Losada, Mayte Y. Li-Gómez, Héctor Cruz-Ramirez, Violeta Alvarez-Venicio, María del Pilar Carreón-Castro, Roberto de J. León-Montiel, Alfred B. U'Ren, "Experimental study on the effects of photon-pair temporal correlations in entangled two-photon absorption," arxiv.org/abs/2101.10987 (2021)

[11] Kristen M. Parzuchowski, Alexander Mikhaylov, Michael D. Mazurek, Ryan N. Wilson, Daniel J. Lum, Thomas Gerrits, Charlies H. Camp Jr, Martin J. Stevens, and Ralph Jimenez. "Setting bounds on two-photon absorption cross-sections in common fluorophores with entangled photon pair absorption." *arXiv preprint arXiv:2008.02664* (2020).

[12] Hong-Bing Fei, Bradley M. Jost, Sandu Popescu, Bahaa EA Saleh, and Malvin C. Teich. "Entanglement-induced two-photon transparency." *Physical review letters* **78**, no. 9 (1997): 1679.

[13] Frank Schlawin, "Entangled photon spectroscopy." *Journal of Physics B: Atomic, Molecular and Optical Physics* **50**, no. 20 (2017): 203001.

[14] Maria Göppert-Mayer, "About elementary acts with two quantum leaps." *Annalen der Physik* **401**, no.3 (1931): 273-294.

[15] Robert W. Boyd, Nonlinear Optics, 3rd Edition, Academic Press (Burlington, 2008), pg 558.

[16] Barak Dayan, Theory of two-photon interactions with broadband down-converted light and entangled photons. *Phys. Rev. A*, **76**: 043813, Oct 2007

[17] Michael Harpham, Özgün Süzer, Chang-Qi Ma, Peter Bäuerle, and Theodore Goodson III. "Thiophene dendrimers as entangled photon sensor materials." *Journal of the American Chemical Society* **131**, no. 3 (2009): 973-979.

[18] R. de J León-Montiel, Jiri Svozilik, L. J. Salazar-Serrano, and Juan P. Torres. "Role of the spectral shape of quantum correlations in two-photon virtual-state spectroscopy." *New Journal of Physics* **15**, no. 5 (2013): 053023.

[19] Michael G. Raymer, Tiemo Landes, Markus Allgaier, Sofiane Merkouche, Brian J. Smith, and Andrew H. Marcus. "Two-photon absorption of time-frequency-entangled photon pairs by molecules: the roles of photon-number correlations and spectral correlations." *arXiv preprint arXiv:2012.05375* (2020).

[20] So-Young Baek and Yoon-Ho Kim. "Spectral properties of entangled photon pairs generated via frequency-degenerate type-I spontaneous parametric down-conversion." *Physical Review A* **77**, no. 4 (2008): 043807.

[21] Warren P. Grice and Ian A. Walmsley. "Spectral information and distinguishability in type-II down-conversion with a broadband pump." *Physical Review A* **56**, no. 2 (1997): 1627.

[22] H. Huang and J. H. Eberly. "Correlations and one-quantum pulse shapes in photon pair generation." *Journal of Modern Optics* **40**, no. 5 (1993): 915-930.

[23] Warren P. Grice, Alfred B. U'Ren, and Ian A. Walmsley. "Eliminating frequency and space-time correlations in multiphoton states." *Physical Review A* **64**, no. 6 (2001): 063815.

[24] S. J. van Enk, (2017). Time-dependent spectrum of a single photon and its positive-operator-valued measure. Physical Review A, **96** (3), 033834.

[25] Michael G. Raymer and Konrad Banaszek. "Time-frequency optical filtering: efficiency vs. temporal-mode discrimination in incoherent and coherent implementations." *Optics Express* **28**, no. 22 (2020): 32819-32836.